\documentclass[fleqn,12pt]{article}

\usepackage[latin2]{inputenc}
\usepackage{t1enc}
\usepackage{amssymb}
\usepackage{amsmath}
\usepackage[mathscr]{eucal}
\usepackage{amsthm}
\usepackage{array}
\usepackage{mathbbol}
\usepackage{color}
\usepackage[pdftex]{graphicx}

\theoremstyle{definition} 
\newtheorem{defin}{Definition}[section]
\newtheorem{thm}[defin]{Theorem}
\newtheorem{rem}[defin]{Remark}
\newtheorem{ex}[defin]{Example}
\newtheorem{cor}[defin]{Corollary}

\newtheorem{lemma}[defin]{Lemma}

\def\vfi{\varphi}
\def\hil{{\mathcal H}}
\def\kil{{\mathcal K}}

\def\B{{\mathcal B}}

\def\X{{\mathcal X}}

\def\S{{\mathcal S}}
\def\M{\mathcal{M}}

\def\half{\frac{1}{2}}
\def\iff{\Longleftrightarrow}

\def\ep{\varepsilon}
\def\N{\mathbb{N}}
\def\iC{\mathbb{C}}
\def\R{\mathbb{R}}

\def\bz{\left(}
\def\jz{\right)}

\def\inv{^{-1}}
\def\kii{\emph}
\def\kiii{}

\def\egy{\mathbf 1}

\def\E{\mathrm{E}}

\def\emax{P_{e,\mathrm{max}}}
\def\eav{P_{e}}

\newcommand{\ki}{\emph}

\newcommand{\s}{\mbox{ }}
\newcommand{\ds}{\mbox{ }\mbox{ }}

\newcommand{\inner}[2]{\langle #1 , #2\rangle}

\newcommand{\vecc}[1]{\underline{#1}}

\newcommand{\diad}[2]{|#1\rangle\langle #2|}
\newcommand{\pr}[1]{\diad{#1}{#1}}

\newcommand{\sr}[2]{S\bz #1\,||\, #2\jz}

\newcommand{\rsr}[3]{S_{#3}\bz #1\,||\, #2\jz}

\newcommand{\srmax}[2]{S_{\mathrm{max}}\bz #1\,||\, #2\jz}

\newcommand{\derleft}[1]{\partial^{-} #1}

\newcommand{\chbound}[2]{C\bz #1\,||\,#2\jz}

\newcommand{\hmeasure}[3]{H_{#3}\bz #1\,||\,#2\jz}

\newcommand{\floor}[1]{\lfloor #1\rfloor}
\newcommand{\channel}[1]{W_{#1}}
\newcommand{\prob}[1]{p_{#1}}
\newcommand{\psif}[3]{\psi_{#1,#2}(#3)}
\newcommand{\vfif}[3]{\vfi_{#1,#2}(#3)}
\newcommand{\hvfif}[3]{\hat\vfi_{#1,#2}(#3)}

\DeclareMathOperator{\Tr}{Tr}
\DeclareMathOperator{\supp}{supp}

\DeclareMathOperator{\ran}{ran}

\begin{document}

\centerline{\huge Generalized relative entropies and}
\medskip

 \centerline{\huge  the capacity of classical-quantum channels}
\bigskip
\s

\bigskip

 \centerline{\large  
 Mil\'an Mosonyi\footnote{Electronic mail: milan.mosonyi@gmail.com}}
   \medskip
 
 \centerline{\textit{Graduate School of Information Sciences, Tohoku University}}
 
 \centerline{\textit{Aoba-ku, Sendai 980-8579, Japan}}
 \bigskip
 
 \centerline{\large Nilanjana Datta\footnote{Electronic mail: N.Datta@statslab.cam.ac.uk}}
 \medskip
 
 \centerline{\textit{Statistical Laboratory, University of Cambridge}}
 
 \centerline{\textit{Wilberforce Road, Cambridge, CB3 0WB, UK}}
 \bigskip
 \s
 
 \medskip

\begin{abstract}
We provide lower and upper bounds on the information transmission capacity of one single use of a classical-quantum channel. 
The lower bound is expressed in terms of the Hoeffding capacity, that we define similarly to the Holevo capacity, but replacing the relative entropy with the Hoeffding distance. Similarly, our upper bound is in terms of a quantity obtained by replacing the relative entropy with the recently introduced max-relative entropy in the definition of the divergence radius of a channel.
\end{abstract}

\medskip

\medskip

\noindent\textit{Comments:}
The following article appeared in J.~Math.~Phys.~\textbf{50}, 072104 (2009) and may be found at
http://link.aip.org/link/?JMP/50/072104.
Copyright (2009) American Institute of Physics. This article may be downloaded for personal use only. Any other use requires prior permission of the author and the American Institute of Physics.
\medskip

\section{Introduction}

The classical channel coding theorem states that the maximum amount of information that can be transmitted through a noisy classical communication channel (asymptotically, per channel use) is equal to the maximum amount of mutual information that can be created between the input and the output of the channel, where the mutual information is measured as the relative entropy distance of a joint distribution from the product of its marginals. 
The celebrated Holevo-Schumacher-Westmoreland (HSW) theorem \cite{Holevo3,SW} states that the same holds for the classical information carrying capacity  
of a quantum channel (or equivalently, the capacity of a classical-quantum channel), for the case of product state inputs and collective measurements on the outputs. 
The capacity is evaluated in the scenario in which the channel is considered to be used an asymptotically large number of times and under the condition that the probability of error in decoding the output, vanishes asymptotically in the number of uses of the channel. 
Moreover, it is assumed that the channel is memoryless, i.e., there is no correlation in the noise of the channel acting on successive input states. 
In real-world applications, however, a channel can only be used finitely many times and
the assumption of the channel being memoryless is not always justifiable, either.
Therefore, it is important to evaluate the optimal rate of information transmission for a finite number of uses of a channel.

In this paper we focus on transmission of classical information through a 
single use of a quantum channel, which can itself correspond to a finite 
number of uses of a channel with arbitrarily correlated noise.
For a general quantum channel, it
is not possible to achieve zero probability of error on a single use. So in this case the capacity is evaluated under the constraint that the probability of error stays below some given threshold $\varepsilon >0$. We hence refer to it as the one-shot $\varepsilon$-capacity of the classical-quantum channel. In this paper we find bounds on this capacity in terms of quantities derived from {\em{generalized relative entropies}}, namely the {\em{Hoeffding distance}} and the {\em{max-relative entropy}}. 

Our main result, Theorem \ref{thm:capacity}, shows that one can find a lower bound on the one-shot capacity of a classical-quantum channel in terms of
its Hoeffding capacity, which is defined in the same way as the Holevo capacity, but with the relative entropy replaced with a Hoeffding distance in its definition.
The main idea of the proof is a combination of the quantum random coding argument of \cite{HN} and a fundamental inequality of hypothesis testing \cite[Theorem 1]{Aud}. 
It is worth noting that hypothesis testing and channel coding are closely related to each other, and hypothesis testing results were already used to obtain coding theorems for classical-quantum channels e.g.~in \cite{Hayashi,HN,ON2}. 
As an application of these techniques, we also show in Theorem \ref{thm:capacity3} a 
lower bound on the exponential capacity of a classical-quantum channel, defined as the optimal asymptotic transmission rate under the constraint that the error probabilities vanish with a given exponential speed.

A geometric interpetation of the asymptotic channel capacity was given in \cite{OPW} (see also \cite{Csiszar} for classical channels), where it was shown that the Holevo capacity of a channel is equal to the divergence radius of its image, as measured by the relative entropy.
In Theorem \ref{thm:upper bound} we show an upper bound on the one-shot capacity of a classical-quantum channel in terms of the
 divergence radius of its image, as measured by the max-relative entropy.

The paper is organized as follows: In Sections \ref{sec:relentr} and \ref{sec:maxrelentr} we introduce the various generalized relative entropies used in the paper, and Section \ref{sec:capacity} is devoted to a brief overview of channel coding and various notions of channel capacities.
In Sections \ref{sec:lower bound} and \ref{sec:upper bound} we prove our lower and upper bounds on the one-shot capacities.
To keep the presentation reasonably compact, we have moved some of the arguments and examples into four separate Appendices.

\section{Preliminaries}\label{sec:Preliminaries}

\subsection{R\'enyi relative entropies and related quantities}\label{sec:relentr}

For a finite dimensional Hilbert space $\hil$, let $\S(\hil)$ denote the set of density operators on $\hil$, and define
\begin{equation*}
\psi:\,\S(\hil)\times\S(\hil)\times\R\to\R,\ds\ds\ds \psi:\,
(\rho,\sigma,t)\mapsto\psif{\rho}{\sigma}{t}:=\log\Tr\rho^t\sigma^{1-t}.
\end{equation*}
(Note that we use the convention $\log 0:=-\infty$ and $0^t:=0,\,t\in\R$. By the latter, powers of a positive semidefinite operator are defined only on its support; in particular, $\rho^0$ stands for the support projection of $\rho$.)
For density operators $\rho,\sigma\in\S(\hil)$, their \ki{R\'enyi relative entropy} of order $t\in [0,1)$ is defined as
\begin{equation*}
\rsr{\rho}{\sigma}{t}:=\frac{1}{t-1}\psif{\rho}{\sigma}{t}=\frac{1}{t-1}\log\Tr \rho^{t} \sigma^{1-t}\,.
\end{equation*}
One can easily see that 
\begin{equation*}
\rsr{\rho}{\sigma}{1}:=\lim_{t\nearrow 1}\rsr{\rho}{\sigma}{t}=\sr{\rho}{\sigma}:=\begin{cases}
\Tr\rho\bz\log\rho-\log\sigma\jz,&\supp\rho\le\supp\sigma,\\
+\infty,&\text{otherwise},
\end{cases}
\end{equation*}
where $\sr{\rho}{\sigma}$ denotes the usual \ki{relative entropy} of $\rho$ and $\sigma$. 

The \ki{Hoeffding distances} of $\rho$ and $\sigma$ are obtained from the R\'enyi relative entropies as
\begin{equation}\label{def:Hoeffding}
\hmeasure{\rho}{\sigma}{r}:=
\sup_{0\le t<1}\frac{-tr-\psif{\rho}{\sigma}{t}}{1-t}=
\sup_{0\le t<1}\left\{\rsr{\rho}{\sigma}{t}-\frac{tr}{1-t}\right\}
\end{equation}
for every $r\ge 0$.
Note that $t\mapsto\rsr{\rho}{\sigma}{t}$ is monotonic increasing 
\cite[Lemma 8]{BD}, and hence,
$\hmeasure{\rho}{\sigma}{0}=\lim_{t\nearrow 1}\rsr{\rho}{\sigma}{t}=\sr{\rho}{\sigma}$. 
It is also clear from the definition that $r\mapsto\hmeasure{\rho}{\sigma}{r}$ is monotonic decreasing, and one can easily see that
\begin{equation}\label{ineq:Hoeffding-relentr}
\rsr{\rho}{\sigma}{0}=\hmeasure{\rho}{\sigma}{\infty}\le
\hmeasure{\rho}{\sigma}{r}\le \hmeasure{\rho}{\sigma}{0}=\sr{\rho}{\sigma},\ds\ds\ds r\ge 0,
\end{equation}
where $\hmeasure{\rho}{\sigma}{\infty}:=\lim_{r\to\infty}\hmeasure{\rho}{\sigma}{r}$.
Let
\begin{equation*}
\vfif{\rho}{\sigma}{a}:=\sup_{0\le t\le 1}\{at-\psif{\rho}{\sigma}{t}\},\ds\ds\ds
\hvfif{\rho}{\sigma}{a}:=\sup_{0\le t\le 1}\{a(t-1)-\psif{\rho}{\sigma}{t}\},\ds\ds\ds
a\in\R.
\end{equation*}
Note that for fixed $\rho,\sigma\in\S(\hil)$, the function $t\mapsto\psif{\rho}{\sigma}{t}$ is convex on $\R$, and $a\mapsto\vfif{\rho}{\sigma}{a}$ is its polar function (or Legendre transform) on the interval $[0,1]$. For an analysis of the properties of these functions, see e.g.~\cite{HMO2}.
It was also shown in \cite{HMO2} that for fixed $\rho$ and $\sigma$ and each $r\ge-\psif{\rho}{\sigma}{1}$, there exists a unique $a_r\le \derleft{\psi_{\rho,\sigma}}(1)$ (the left derivative of $\psi_{\rho,\sigma}$ at $1$) such that $\hvfif{\rho}{\sigma}{a_r}=r$, and
\begin{equation}\label{formula:Hoeffding}
\hmeasure{\rho}{\sigma}{r}=\vfif{\rho}{\sigma}{a_r},\ds\ds\text{i.e.,}\ds\ds
\hmeasure{\rho}{\sigma}{r}=\bz\vfi_{\rho,\sigma}\circ\hat\vfi_{\rho,\sigma}\inv\jz(r),\ds\ds r\ge -\psif{\rho}{\sigma}{1}.
\end{equation}
Note that $t\mapsto\hat\vfi_{\rho,\sigma}(t)$ is strictly monotonically decreasing on the interval $(-\infty, \derleft{\psi_{\rho,\sigma}}(1)]$, and $\hat\vfi_{\rho,\sigma}\inv$ 
denotes its inverse on this interval. Since both $\vfi_{\rho,\sigma}$ and $\hat\vfi_{\rho,\sigma}$ are continuous, \eqref{formula:Hoeffding} yields
\begin{equation}\label{limit}
\lim_{r\searrow 0}\hmeasure{\rho}{\sigma}{r}=\hmeasure{\rho}{\sigma}{0}=\sr{\rho}{\sigma}.
\end{equation}

Finally, the \ki{Chernoff distance} of $\rho,\sigma\in\S(\hil)$ is defined from the $\psi$ function as
\begin{equation*}
\chbound{\rho}{\sigma}:=\vfif{\rho}{\sigma}{0}=-\min_{0\le t\le 1}\psif{\rho}{\sigma}{t}.
\end{equation*}
One can easily see that the Chernoff distance also falls between $S_0$ and $S_1$, i.e.,
\begin{equation*}
\rsr{\rho}{\sigma}{0}\le\chbound{\rho}{\sigma}\le\sr{\rho}{\sigma}.
\end{equation*}

The R\'enyi relative entropies, the Chernoff distance and the Hoeffding distances are all non-negative
and hence can be considered as generalized distances between states (though they are not symmetric in their variables, except for the Chernoff distance, and do not satisfy the triangle inequality). The relative entropy and the Chernoff distance are also strictly positive, unless the two states are equal.
Due to Lieb's concavity theorem \cite{Lieb} and Uhlmann's method \cite{Uhlmann}, all these quantities are jointly convex in the variables $(\rho,\sigma)$ and
monotonic decreasing under stochastic (i.e., completely positive and trace-preserving) maps acting simultaneously on $\rho$ and $\sigma$ (see \cite{Petz} for an alternative proof). Finally, all these quantities emerge naturally as the optimal decay rates of certain error probabilities in asymptotic hypothesis testing problems; see, e.g.~\cite{Aud,ANSzV,BS,Hayashi,HMO,HMO2,HP,HP2,MHOF,M,Nagaoka,NSz,ON}.
 
\subsection{The max-relative entropy}\label{sec:maxrelentr}

Following \cite{Datta}, we define 
the \ki{max-relative entropy} of states $\rho$ and $\sigma$ as
\begin{equation*}
\srmax{\rho}{\sigma}:=\log\inf\{\lambda\,:\,\rho\le\lambda\sigma\}=\inf\{\gamma\,:\,
\rho\le 2^{\gamma}\sigma\}.
\end{equation*}
(Note that our notation here differs from that of \cite{Datta}, where the max-relative entropy of states
$\rho$ and $\sigma$ was denoted by the symbol $D_{\max}(\rho||\sigma)$).
One can easily see that for commuting $\rho$ and $\sigma$ with $\supp\rho\le\supp\sigma$, the max-relative entropy coincides with the R\'enyi relative entropy of parameter $\infty$, defined as $\rsr{\rho}{\sigma}{\infty}:=\lim_{t\to\infty}\frac{1}{t-1}\log\Tr\rho^t\sigma^{1-t}$.
The truly quantum case, however, is different, and the max-relative entropy turns out to be an independent quantity; see e.g.~Example \ref{count1}. In the general case, $S_{\max}$ and $S_{\infty}$ are related as 
\begin{equation*}
\srmax{\rho}{\sigma}\le\rsr{\rho}{\sigma}{\infty}
\end{equation*}
whenever $\supp\rho\le\supp\sigma$ \cite{Colbeck}.
One can see from the definition that 
\begin{equation}\label{ineq:relente-max}
\sr{\rho}{\sigma}\le\srmax{\rho}{\sigma}
\end{equation}
for all states $\rho,\sigma$.
In particular, the max-relative entropy is also strictly positive (unless the two states are equal). It also follows 
easily from the definition that the max-relative entropy is monotonic decreasing under arbitrary positive (not 
necessarily stochastic) maps acting simultaneously on $\rho$ and $\sigma$.
These and other properties of the max- relative entropy were discussed in \cite{Datta}. On the other hand, the max-relative entropy 
is not jointly convex in its variables in general; see e.g.~Example \ref{count2}. 

The max-relative entropy is also related to the optimal performance in a state discrimination problem, as it was shown recently in \cite{KRS}. Consider a multiple state discrimination problem where the hypotheses $\rho_1,\ldots,\rho_M$ to discriminate are states on some Hilbert space $\hil$. The optimal average success probability is given as $P_{s}^*:=\sup_{\{E_1,\ldots,E_m\}}(1/M)\sum_{k=1}^M\Tr\rho_k E_k$, where the supremum is taken over all positive operator valued measures (POVM) $\{E_1,\ldots,E_M\},\,0\le E_k\le I,\,\sum_{k=1}^M E_k=I$. Theorem 1 in \cite{KRS} yields that
\begin{equation}\label{thm:KRS}
P_s^*=\frac{1}{M}\inf_{\sigma\in\S(\hil)}\max_{1\le k\le M}2^{\srmax{\rho_k}{\sigma}}.
\end{equation}
Since our formulation here is slightly different from that of \cite{KRS}, for readers' convenience we give a brief sketch of the proof of \eqref{thm:KRS} in \ref{Renner}.

\subsection{Capacities of classical-quantum channels}\label{sec:capacity}

By a \ki{classical-quantum communication channel} (or simply a \ki{channel}) we mean a  triple $(\X,\hil,W)$, where 
$\X$  is a set, $\hil$ is a Hilbert space and $W$ maps elements of $\X$ into density 
operators on $\hil$. If no confusion arises, we will denote the channel simply by $W$. 
Elements of $\X$ are the possible inputs for the channel and $\ran W$ is the set of the 
possible outputs, which we will also call the \ki{image} of the channel. The channel is 
\ki{classical} if its image is a commutative subset of $\B(\hil)$. Note that the standard 
definition of a quantum channel is recovered by choosing $\X$ to be the state space 
$\S(\hil_{in})$ of some Hilbert space $\hil_{in}$ and $W$ to be a completely positive 
trace-preserving linear map from $\B(\hil_{in})$ to $\B(\hil)$. 

In order to use the channel for transmitting (classical) messages, one has to assign a 
codeword to each message, which is an element in the input set $\X$. After the message is 
transmitted through the channel, the receiver has to decide which message was sent. If the 
receiver knows the codewords and how the channel acts on them, then his task is to perform 
state discrimination on the possible outcomes of the channel. 
We say that a triple $(M,\vfi,\E)$ is an \ki{$M$-code} if $\vfi$ is a function from 
$\{1,\ldots,M\}$ to $\X$ (the \ki{encoding}) and $E$ is a function from $\{1,\ldots,M\}$ to 
$\B(\hil)$ (the \ki{decoding}) such that $E_k\ge 0,\,k=1,\ldots,M$, and $\sum_{k=1}^M 
E_k\le I$. Here, $1,\ldots,M$ are the labels of the messages the sender would like to transmit 
through the channel, $\vfi_1,\ldots,\vfi_M$ are the codewords, and $E_1,\ldots, E_M$ are 
the POVM operators to discriminate the states $\channel{\vfi_1},\ldots,\channel{\vfi_M}$ at the output 
of the channel. The average error probability of such an $M$-code is
 \begin{equation*}
\eav(M,\vfi,\E):=\frac{1}{M}\sum_{k=1}^M \Tr \channel{\vfi_k}(I-E_k).
 \end{equation*}
For a given $\ep>0$, the \ki{one-shot $\ep$-capacity} of the channel is the maximum number of bits that can be transmitted through the channel with error probability at most $\ep$: 
\begin{equation*}
C_{\ep}(W):=\sup\{\log M\,:\,\text{there exists an $M$-code with }\eav(M,\vfi,\E)\le\ep\}.
\end{equation*}
Here, the base of the logarithm is chosen to be $2$. 
Note that one could also define the $\ep$-capacity using the maximum error probability $\emax(M,\vfi,\E):=\max_{1\le k\le M}\Tr \channel{\vfi_k}(I-E_k)$ instead. This capacity $C_{\ep,\max}(W)$ is related to $C_{\ep}(W)$ as
$C_{\ep/2}(W)-1\le C_{\ep,\max}(W)\le C_{\ep}(W)$, where the second inequality is obvious and the first one follows by "throwing away the worst half of the codewords" (see e.g. \cite[p.~204]{CT}).

Consider now  the $n$th product extension of the channel $W$, defined as
\begin{equation*}
W^{(n)}:\,\X^n\to\S(\hil^{\otimes n}),\ds\ds\ds
W^{(n)}(x_1,\ldots,x_n):=W(x_1)\otimes\ldots\otimes W(x_n).
\end{equation*}
Note that if $W$ is a quantum channel with $\X=\S(\hil_{\mathrm{in}})$ then
\begin{equation*}
W^{(n)}(\rho_1,\ldots,\rho_n)=W^{\otimes n}\bz\rho_1\otimes\ldots\otimes\rho_n\jz,\ds\ds\ds
\rho_1,\ldots,\rho_n\in\S(\hil_{\mathrm{in}}).
\end{equation*}
Hence, this formulation only allows product encoding, while entangled measurement is allowed in the decoding.
The \ki{asymptotic $\ep$-capacity} of $W$ is defined as
\begin{equation*}
\overline{C}_{\ep}(W):=\sup\left\{\liminf_n\frac{1}{n}\log M^{(n)}\,:\, \limsup_n P_{e}(M^{(n)},\vfi^{(n)},E^{(n)})\le\ep\right\},
\end{equation*}
where the supremum is taken over sequences of codes $(M^{(n)},\vfi^{(n)},E^{(n)})$, satisfying the indicated criterion. One can easily see that 
\begin{equation}\label{ineq:capacity}
\liminf_n\frac{1}{n}C_{\ep}(W^{(n)})\le\overline{C}_{\ep}(W)\le \overline{C}_{\ep'}(W)\le\liminf_n\frac{1}{n}C_{\ep''}(W^{(n)})
\end{equation}
for any $0\le \ep\le\ep'<\ep''$. 
One can also define a stronger notion of asymptotic capacity by requiring that the error probabilities vanish with a given exponential speed:
\begin{equation*}
\overline{C}_{r}^{\exp}(W):=\sup\left\{\liminf\frac{1}{n}\log M^{(n)}\,:\, \limsup_n\frac{1}{n}\log P_{e}(M^{(n)},\vfi^{(n)},E^{(n)})<-r\right\}.
\end{equation*}
Obviously, $r\mapsto \overline{C}_{r}^{\exp}$ is monotonic decreasing, and $\overline{C}_{r}^{\exp}\le \overline{C}_{0}^{\exp}\le \overline{C}_{0}\le\overline{C}_{\ep}$
for any $0\le r,\ep$.

Let $\M_f(\X)$ denote the set of finitely supported probability measures on $\X$, and define
$\kil:=l^2(\X)$, the $L^2$-space on $\X$ with respect to the counting measure.
For each $x\in\X$, define the rank-one projection $\delta_x:=\pr{\egy_{\{x\}}}$, where $\egy_{\{x\}}$ is the characteristic function of the one-point set $\{x\}$. For a finitely supported probability measure $p$ on $\X$, let 
\begin{equation}\label{rpqp}
R_p:=\sum_{x\in\X}\prob{x}\delta_x\otimes \channel{x},\ds\ds\ds
Q_p:=\bz\sum_{x\in\X}\prob{x}\delta_x\jz\otimes E_p(W),
\end{equation}
where $E_p(W):=\sum_x \prob{x} \channel{x}$.
Obviously, $R_p$ and $Q_p$ are density operators on $\kil\otimes\hil$, and 
$Q_p$ is the product of the marginals of $R_p$. Hence, $\sr{R_p}{Q_p}$ is the mutual information in the bipartite classical-quantum state $R_p$, defined as its distance from the product of its marginals, where the distance is measured by the relative entropy.
The Holevo-Schumacher-Westmoreland theorem \cite{Holevo3,SW} states that 
\begin{equation*}
\overline{C}_{0}(W)=\chi^*(W):=\sup_{p\in\M_f(\X)}\sr{R_p}{Q_p}.
\end{equation*}
The quantity $\chi^*(W)$ is called the \ki{Holevo capacity} of $W$.

\section{Hoeffding capacities and lower bounds}\label{sec:lower bound}

It is a natural idea to measure the amount of correlations in a bipartite state as its distance from the 
product of its marginals, and the channel coding theorem selects the relative entropy as the 
right measure of distance. One may, however, define the amount of correlations 
using some generalized relative entropy $D(.\,||\,.)$, and define the corresponding version of the Holevo capacity as $\chi^*_{_D}(W):=\sup_{p\in\M_f(\X)}D(R_p\,||\,Q_p)$. In particular, for a channel $W$ we define its \ki{Hoeffding capacity} with parameter $r\ge 0$ as
\begin{equation*}
\chi^*_{_r}(W):=\chi^*_{_{H_r}}(W):=\sup_{p\in\M_f(\X)}\hmeasure{R_p}{Q_p}{r},
\end{equation*}
where $R_p$ and $Q_p$ are as in \eqref{rpqp}.
Note that if $D$ is the relative entropy then for any $p\in\M_f(\X)$,
\begin{equation*}
\sr{R_p}{Q_p}=\sum_x\prob{x}\sr{\channel{x}}{E_p(W)}=S(E_p(W))-\sum_x\prob{x}S(\channel{x}),
\end{equation*}
where $S(\rho):=-\sr{\rho}{I}$ is the von Neumann entropy of a state $\rho$. These identities
are specific to the relative entropy and do not hold for a general $D$. However, if $D$ is jointly convex in its variables and invariant under adding an ancilla then
\begin{equation*}
D(R_p\,||\,Q_p)\le \sum_x\prob{x} D\bz\delta_x\otimes \channel{x}\,||\,\delta_x\otimes E_p(W)\jz
=\sum_x\prob{x} D\bz W_x\,||\,E_p(W)\jz.
\end{equation*}
This holds, for instance, for the R\'enyi relative entropies with parameter $t\in[0,1]$, the Hoeffding distances and the Chernoff distance.

Our main goal in this section is to give lower bounds on the one-shot capacity and the exponential capacities of a classical-quantum channel in terms of its Hoeffding capacity.
We will make use of the following lemma, which is essentially the same as inequality (11) in \cite{Hayashi}. For readers' convenience, we give a detailed proof in \ref{proofs}.
\begin{lemma}\label{cor:error bound}
For any $M\in\N$, any $c>0$ and any $p$ finitely supported probability distribution on $\X$, there exists an $M$-code $(M,\vfi,E)$ such that 
\begin{equation*}
P_{e}(M,\vfi,E)\le (1+c)^t(2+c+1/c)^{1-t}(M-1)^{1-t}\Tr R_p^tQ_p^{1-t},\ds\ds\ds 0\le t\le 1.\hfill\qed
\end{equation*}
\end{lemma}

\begin{thm}\label{thm:capacity}
For any channel $W$, the one-shot $\ep$-capacity is lower bounded as
\begin{equation}\label{epcapacity lower bound}
C_{\ep}(W)\ge \chi^*_{\log\bz\frac{1+c}{\ep}\jz}(W)-\log\bz\frac{2+c+1/c}{\ep}\jz
\end{equation}
for any $c>0$.
\end{thm}
\begin{proof}
By Lemma \ref{cor:error bound},
\begin{equation}\label{lower bound}
C_{\ep}(W)\ge \log M
\end{equation}
for any $M$ such that there exist a $p\in\M_f(\X)$, a $c>0$ and a $t\in[0,1)$ such that 
\begin{equation*}
(1+c)^t(2+c+1/c)^{1-t}(M-1)^{1-t}\Tr R_p^tQ_p^{1-t}\le\ep.
\end{equation*}
Rewriting this condition, we get
\begin{align*}
\log(M-1)&\le 
\frac{1}{1-t}\bz t\log(\ep/(1+c))+(1-t)\log\bz\ep/(2+c+1/c)\jz-\log\Tr R_p^tQ_p^{1-t}\jz\\
&\le
 -\log\bz(2+c+1/c)/\ep\jz+\sup_{0\le t<1}\frac{-t\log\bz(1+c)/\ep\jz-\log\Tr R_p^tQ_p^{1-t}}{1-t}\\
&= -\log\bz(2+c+1/c)/\ep\jz+\hmeasure{R_p}{Q_p}{\log\bz(1+c)/\ep\jz},
\end{align*}
from which the statement follows.
\end{proof}

\begin{rem}\label{Remark}
The term $\chi^*_{\log\bz\frac{1+c}{\ep}\jz}(W)-\log\bz\frac{2+c+1/c}{\ep}\jz$ in Theorem \ref{thm:capacity} goes to $-\infty$ as $\ep$ tends to $0$ and hence the lower bound in Theorem \ref{thm:capacity} becomes trivial below some threshold $\ep_0$.
This comes of course as no surprise: even if the asymptotic capacity of the channel is non-zero, it might not be possible to transmit more than one message with arbitrarily small error probability in one single use of the channel; see, e.g., Example \ref{classical depolarizing}.

Note that the Hoeffding distance is monotonically decreasing in its parameter and hence,
with the choice $c=1$ in Theorem \ref{thm:capacity}, we get
\begin{equation}\label{lower bound2}
C_{\ep}(W)\ge\chi^*_{_{\log(2/\ep)}}(W)-\log(4/\ep)\ge \chi^*_{_{\log(4/\ep)}}(W)-\log(4/\ep).
\end{equation}
This lower bound is strictly positive if and only if 
there exists a $p\in\M_f(\X)$ for which 
\begin{equation*}
\hmeasure{R_p}{Q_p}{\log(4/\ep)}-\log(4/\ep)>0.
\end{equation*}
Note that $\supp R_p\le\supp Q_p$ and hence $\psi_{_{R_p,Q_p}}(1)=0$, and \eqref{formula:Hoeffding} implies that for any $r\ge 0$,
\begin{equation*}
\hmeasure{R_p}{Q_p}{r}-r=\vfi_{_{R_p,Q_p}}(a_r)-r=\vfi_{_{R_p,Q_p}}(a_r)-\hat\vfi_{_{R_p,Q_p}}(a_r)=a_r=\hat\vfi_{_{R_p,Q_p}}\inv(r).
\end{equation*}
Note that $a_r=0$ is equivalent to 
\begin{equation*}
r=\hat\vfi_{_{R_p,Q_p}}(0)=\vfi_{_{R_p,Q_p}}(0)=\chbound{R_p}{Q_p},
\end{equation*}
the Chernoff information in the classical-quantum state $R_p$. Since $a_r=\hat\vfi_{_{R_p,Q_p}}\inv(r)$, and $\hat\vfi_{_{R_p,Q_p}}$ is monotonically decreasing, we finally get that 
\begin{equation*}
\hmeasure{R_p}{Q_p}{r}-r>0 \iff r<\chbound{R_p}{Q_p}.
\end{equation*}
Hence, the lower bound in \eqref{lower bound2} is strictly positive if and only if
\begin{equation*}
\log(4/\ep)<\chi^*_{_C}(W),\ds\ds\text{or equivalently,}\ds\ds
2^{2-\chi^*_{_C}(W)}<\ep,
\end{equation*}
where $\chi^*_{_C}(W):=\sup_{p\in\M_f(\X)}\chbound{R_p}{Q_p}$ is the \ki{Chernoff capacity} of the channel.
\hfill\qed
\end{rem}

\begin{rem}
Example \ref{classical depolarizing} shows that for any $\ep\in [0,1/2)$ and any $K>0$ there exists a channel $W$ such that $C_{\ep}(W)=0$ while $\chi^*(W)>K$. This shows that there exists no function $f:\,\R_{+}\to\R_{+}$ for which 
$C_{\ep}(W)\ge \chi^*(W)-f(\ep)$ would hold for every channel. Hence, we cannot have a lower bound similar to \eqref{epcapacity lower bound} with $\chi^*(W)$ in place of the Hoeffding capacity.
\hfill\qed
\end{rem}

Theorem \ref{thm:capacity} yields immediately the following:
\begin{cor}\label{cor:nshot}
For any channel $W$, any $c>0$, any $\ep>0$ and any $n\in\N$, the capacity per channel use for $n$ uses of the channel is lower bounded as
\begin{equation*}
\frac{1}{n}C_{\ep}\bz W^{(n)}\jz\ge 
\chi^*_{_{\frac{1}{n}\log((1+c)/\ep)}}(W)-\frac{1}{n}\log((2+c+1/c)/\ep).
\end{equation*}
\end{cor}
\begin{proof}
Note that for any $p\in\M_f(\X)$ and any $n\in\N$, 
\begin{equation*}
R_{p^{\otimes n}}=R_p^{\otimes n},\ds
Q_{p^{\otimes n}}=Q_p^{\otimes n}\ds
\text{and}\ds
\hmeasure{R_{p^{\otimes n}}}{Q_{p{\otimes n}}}{r}
=n\hmeasure{R_p}{Q_p}{r/n}
\end{equation*}
for any $r\ge 0$.
By Theorem \ref{thm:capacity},
\begin{eqnarray*}
C_{\ep}\bz W^{(n)}\jz&\ge& 
\hmeasure{R_{p^{\otimes n}}}{Q_{p^{\otimes n}}}{\log((1+c)/\ep)}-\log((2+c+1/c)/\ep)\\
&=&
n\left[\hmeasure{R_p}{Q_p}{\frac{1}{n}\log((1+c)/\ep)}-\frac{\log((2+c+1/c)/\ep)}{n}\right],
\end{eqnarray*}
from which the statement follows.
\end{proof}

Theorem \ref{thm:capacity} only provides a lower bound on the one-shot $\ep$-capacity of a channel. However, it is asymptotically sharp in the sense that the lower bound of the HSW theorem can be recovered from it. Indeed,
by Corollary \ref{cor:nshot}, the first inequality in \eqref{ineq:capacity}, and by \eqref{limit},
\begin{eqnarray*}
\overline{C}_{\ep}(W)&\ge&\liminf_n\frac{1}{n}C_{\ep}(W^{(n)})\\
&\ge& \lim_n \left[\hmeasure{R_p}{Q_p}{\frac{1}{n}\log((1+c)/\ep)}-\frac{\log((2+c+1/c)/\ep)}{n}\right]\\
&=&
\hmeasure{R_p}{Q_p}{0}=\sr{R_p}{Q_p},
\end{eqnarray*}
from which $\overline{C}_{\ep}(W)\ge \chi^*(W)$ follows for all $\ep>0$. Considering 
a sequence $\ep_n\to 0$, one also gets $\overline{C}_{0}(W)\ge \chi^*(W)$.

It is known that for rates below the capacity $\chi^*(W)$, one can find a sequence of codes for which the error probabilities vanish with an exponential speed, and hence 
$\overline{C}_{0}^{\exp}(W)\ge \chi^*(W)$. 
One can use Lemma \ref{cor:error bound} to give a lower bound on the exponential capacities, that yields the above lower bound as a special case:
\begin{thm}\label{thm:capacity3}
For any channel $W$ and $r\ge 0$,
\begin{equation*}
\overline{C}_{r}^{\exp}(W)\ge \chi_{_r}^*(W)-r.
\end{equation*}
\end{thm}
\begin{proof}
The statement is trivial if $\chi_{_r}^*(W)-r\le 0$, hence for the rest we assume it to be strictly positive. Let $0<R<\chi_{_r}^*(W)-r$. By definition, there exists a $p\in\M_f(\X)$ such that 
$R<\hmeasure{R_p}{Q_p}{r}-r$. 
By the definition of the Hoeffding distances,
there exists a $t\in[0,1)$ such that 
\begin{equation*}
R<-r-\frac{tr}{1-t}+\rsr{R_p}{Q_p}{t}=\frac{1}{1-t}\bz -r-\log\Tr R_p^t Q_p^{(1-t)}\jz,
\end{equation*}
or equivalently,
\begin{equation*}
2^{(1-t)R}\Tr R_p^tQ_p^{1-t}<2^{-r}.
\end{equation*}
Now, for each $n\in\N$ we can apply
Lemma \ref{cor:error bound} with
the channel being $W^{(n)}$, the probability distribution $p^{\otimes n}$ and 
 $M^{(n)}:=\floor{2^{nR}}$, and get the existence of an $M^{(n)}$-code 
$(M^{(n)},\vfi^{(n)},E^{(n)})$ such that 
\begin{equation*}
P_{e}(M^{(n)},\vfi^{(n)},E^{(n)})\le 4\floor{2^{nR}}^{1-t}\Tr R_{p^{\otimes n}}^t Q_{p^{\otimes n}}^{1-t}.
\end{equation*}
Since $R_{p^{\otimes n}}=\bz R_p\jz^{\otimes n}$ and $Q_{p^{\otimes n}}=\bz Q_p\jz^{\otimes n}$, we finally get
\begin{equation*}
P_{e}(M^{(n)},\vfi^{(n)},E^{(n)})
\le 4\bz 2^{(1-t)R}\Tr R_p^tQ_p^{1-t}\jz^n
<4\cdot 2^{-nr},
\end{equation*}
from which the assertion follows.
\end{proof}
\begin{rem}
As we have seen in Remark \ref{Remark}, the lower bound $\chi^*_{_r}(W)-r$ is strictly positive if and only if $r<\chi^*_{_C}(W)$.
\end{rem}

\section{Divergence radii and an upper bound}\label{sec:upper bound}

The \ki{divergence radius} of a subset $\Sigma\subset\S(\hil)$ with respect to some generalized relative entropy $D$ is defined as
\begin{equation*}
R_D(\Sigma):=\inf_{\sigma\in\S(\hil)}\sup_{\rho\in\Sigma}\{D(\rho\,||\,\sigma)\}.
\end{equation*}
In particular, we denote by $R_{\max}(\Sigma):=R_{S_{\max}}(\Sigma)$ the max-relative entropy radius of $\Sigma$. We have the following:

\begin{thm}\label{thm:upper bound}
For any channel $W$ and $\ep>0$,
\begin{equation*}
C_{\ep}(W)\le R_{\max}(\ran W)-\log(1-\ep).
\end{equation*}
\end{thm}
\begin{proof}
Let $(M,\vfi,E)$ be an $M$-code for which $P_{e}(M,\vfi,E)\le\ep$. By \eqref{thm:KRS},
\begin{eqnarray*}
P_{e}(M,\vfi,E)&\ge &
1-\inf_{\sigma\in\S(\hil)}\max_{1\le k\le M}\frac{1}{M}2^{\srmax{W_{\vfi_k}}{\sigma}}\\
&=&
1-\frac{1}{M}2^{R_{\max}(\{W_{\vfi_k}\})},
\end{eqnarray*}
which yields
\begin{equation*}
\log M<-\log(1-\ep)+R_{\max}(\{W_{\vfi_k}\})\le -\log(1-\ep)+R_{\max}(\ran W),
\end{equation*}
from which the statement follows.
\end{proof}

\begin{rem}
An alternative proof of the above Theorem can be obtained using Lemma 4 in \cite{HN}, which states that for any code $(M,\vfi,E)$, any state $\sigma\in\ran W$ and any $\gamma\in\R$,
\begin{equation*}
P_e(M,\vfi,E)\ge \frac{1}{M}\sum_{k=1}^M\Tr W_{\vfi_{k}}\{2^{\gamma}\sigma-W_{\vfi_{k}}\ge 0\}-\frac{2^{\gamma}}{M}.
\end{equation*}
Choosing therefore $\gamma:=R_{\max}(\ran W)$ and $\sigma$ to be a state where the infimum in the definition of $R_{\max}(\ran W)$ is attained, one obtains
\begin{equation*}
P_e(M,\vfi,E)\ge 1-2^{R_{\max}}(\ran W)/M.
\end{equation*}
Therefore, $P_e(M,\vfi,E)\le \ep$ yields $\log M\le R_{\max}(\ran W)-\log(1-\ep)$.
\end{rem}

\begin{rem}
The additivity of the max-relative entropy on product states yields that $R_{\max}\bz\ran W^{(n)}\jz\le nR_{\max}(W)$ and hence, by Theorem \ref{thm:upper bound},
\begin{equation*}
\overline{C}_{\ep}(W)\le \liminf_n\frac{1}{n} C_{\ep'}(W^{(n)})\le \liminf_n\frac{1}{n}R_{\max}(\ran W^{(n)})\le R_{\max}(\ran W)
\end{equation*}
for any $\ep<\ep'<1$. This upper bound, however, is not optimal in general, as Example \ref{classical depolarizing} shows.
\end{rem}
 
\begin{rem}
As noted before, $R_D(\ran W)=\chi^*_{_D}(W)$ when $D$ is the relative entropy. 
For a general $D$, such an identity does not hold. However, when $D$ is the max-relative entropy, $\srmax{R_p}{Q_p}=\max_{x\in\supp p}\srmax{\channel{x}}{E_p(W)}$ yields
\begin{align*}
R_{\max}(\ran W)&=
\sup_{p\in\M_f(\X)}\inf_{\sigma\in\S(\hil)}\sup_{x\in \supp p}\{\srmax{\channel{x}}{\sigma}\}\\
&\le
\sup_{p\in\M_f(\X)}\max_{x\in \supp p}\{\srmax{\channel{x}}{E_p(W)}\}\\
&=
\sup_{p\in\M_f(\X)}\srmax{R_p}{Q_p}
=
\chi^*_{_{\max}}(W).
\end{align*}
\end{rem}

\section{Conclusion}

We have shown lower bounds on the one-shot capacities and the exponential capacities of a classical-quantum channel in terms of its Hoeffding capacity, and an upper bound in terms of the max-relative entropy radius of its image. While the lower bounds on the one-shot capacities were shown to be asymptotically tight, the same is not known for the upper bounds of Theorem \ref{thm:upper bound}.
It is an open question whether a sensible upper bound can also be found in terms of the Hoeffding capacities. To the best of our knowledge, our lower bound is a new result even for classical channels.

The exponential capacities considered in this paper are in some sense  dual to the well-known notion of the error exponent in channel coding theory. The latter is defined as the optimal exponential decay rate of the error probabilities for sequences of codes with a fixed transmission rate. An upper bound on the error exponent was given in inequality (11) in \cite{Hayashi}, and one can easily verify that the lower bound in our Theorem \ref{thm:capacity3} can actually be derived from that.

In Stein's lemma of hypothesis testing, one is interested in the asymptotic behaviour of the quantities $(1/n)\beta_{\ep}(\rho_n\,||\,\sigma_n)$ for two sequences of states $\{\rho_n\}_{n\in\N}$ and $\{\sigma_n\}_{n\in\N}$ and some $\ep\in(0,1)$, where
$\beta_{\ep}(\rho_n\,||\,\sigma_n):=\inf\{\log\Tr\sigma_n A\,:\,0\le A\le I,\s \Tr \rho_n(I-A)\le \ep\}$ is the logarithm of the optimal error probability of the second kind under the constraint that the error probability of the first kind stays below $\ep$. When $\rho_n$ and $\sigma_n$ are the $n$th i.i.d.~extensions of the states $\rho_1$ and $\sigma_1$, respectively, then $\lim_{n\to\infty}(1/n)\beta_{\ep}(\rho_n\,||\,\sigma_n)=-\sr{\rho_1}{\sigma_1}$ for any $\ep\in(0,1)$ \cite{HP,ON}. This result 
provides an operational interpretation of the relative entropy, and was used in \cite{ON2} to give an alternative proof for  
the achievability part of the HSW theorem, namely that $\overline{C}_0(W)\ge \sup_{p\in\M_f(\X)}\sr{R_p}{Q_p}$. 
Recently, upper and lower bounds on the one-shot capacities of classical \cite{WCR} and classical-quantum channels \cite{WR} were obtained in terms of the quantities $\beta_{\ep}(R_p\,||\,Q_p)$. These results 
refine the connection between channel coding and hypothesis testing by 
establishing a connection between the operational quantities of the two theories.
At the moment it is not clear how the lower bounds of \cite{WCR,WR} and our lower bound are related to each other.

\section*{Acknowledgments}
The research leading to these results has received funding from the European Community's Seventh Framework Programme (FP7/2007-2013) under grant agreement no.~213681.
MM received partial funding from the Grant-in-Aid for JSPS
Fellows 18\,$\cdot$\,06916 and the Hungarian Research Grant
OTKA T068258. The authors would like to thank Prof.~Fumio Hiai for helpful comments on the manuscript.
ND would like to thank Fernando Brandao, Andreas Winter and Keiji Matsumoto for earlier discussions. 
MM would like to thank the Statistical Laboratory, University of Cambridge, for kind hospitality.
This research was started during his visit there, and it was completed while he was a Junior Research Fellow at the Erwin Schr\"odinger Institute in Vienna.

\appendix
\swapnumbers

\def\thesection{Appendix \Alph{section}} 
\section{} \label{counterexample}

\def\thesection{\Alph{section}} 

\begin{ex}\label{count1}
Let $0<a<1/2$ and define density operators 
\begin{equation*}
\rho:=\frac{1}{2}\begin{bmatrix} 1 &1\\ 1 & 1\end{bmatrix}\ds\ds\text{and}\ds\ds
\sigma:=\begin{bmatrix} a & 0\\ 0 & 1-a\end{bmatrix}
\end{equation*}
on $\hil:=\iC^2$. 
One can easily see that $\lambda\sigma-\rho\ge 0$ if and only if
$\lambda\ge \frac{1}{2a(1-a)}$, and hence
\begin{equation*}
\srmax{\rho}{\sigma}=\log\frac{1}{2a(1-a)}.
\end{equation*}
On the other hand, a straightforward computation yields
\begin{equation*}
\rsr{\rho}{\sigma}{\infty}=-\log\min\{a,1-a\}=\log\frac{1}{a}.
\end{equation*}
By assumption, $2(1-a)>1$, and hence,
\begin{equation*}
\srmax{\rho}{\sigma}<\rsr{\rho}{\sigma}{\infty}.
\end{equation*}
\end{ex}

\begin{ex}\label{count2}
Let $\rho_k,\sigma_k,\,k=1,\ldots,r$ be density operators on a Hilbert space $\hil$ such that
$\supp\rho_k\le \supp\sigma_k$ for all $k$. Let $\delta_k,\,k=1,\ldots,r$, be a set of orthogonal rank-one projections in some auxiliary Hilbert space $\kil$, and let $p_1,\ldots,p_r$ be strictly positive convex weights. Then,
\begin{equation*}
\srmax{\sum_k \prob{k}\delta_k\otimes\rho_k}{\sum_k \prob{k}\delta_k\otimes\sigma_k}=
\max_k\srmax{\rho_k}{\sigma_k}\nleq\sum_k \prob{k} \srmax{\rho_k}{\sigma_k}
\end{equation*}
unless $\srmax{\rho_k}{\sigma_k}$ is the same for all $k$.
\end{ex}

 \def\thesection{Appendix \Alph{section}} 
 \section{} \label{Renner}
 \def\thesection{\Alph{section}}
 
Let $\egy_{\{1\}},\ldots,\egy_{\{M\}}$ be the standard basis of $\iC^M$, and define $\tilde\rho:=\sum_{k=1}^r (1/M)\pr{\egy_{\{k\}}}\otimes\rho_k$.
The optimal success probability can be expressed as
\begin{equation*}
P_{s}^*=\sup_{\{E_1,\ldots,E_m\}}\Tr \tilde\rho
\sum_{k=1}^r\pr{\egy_{\{k\}}}\otimes E_k=\sup_{\substack{0\le E\in\B(\iC^M\otimes\hil),\\ \Tr_{\iC^M}E=I_{\hil}}}\Tr\tilde\rho E,
\end{equation*}
where the first supremum is taken over all POVM $\{E_1,\ldots,E_M\}$.
Using the duality theorem of linear programming, it was shown in \cite[Lemma 4]{KRS} (see also formula III.15 in \cite{YKL}), that the right-hand side of the above formula is equal to 
\begin{eqnarray*}
\inf\{\Tr B\,:\,B\ge 0,\,\tilde\rho\le I_{\iC^M}\otimes B\}&=&
\inf\{\Tr B\,:\,B\ge 0,\,\rho_k\le MB,\,k=1,\ldots,M\}\\
&=&
\frac{1}{M}\inf\{\Tr B\,:\,B\ge 0,\,\rho_k\le B,\,k=1,\ldots,M\}.
\end{eqnarray*}
Replacing the infimum over $B$ with infima over $\sigma:=(1/\Tr B) B$, and $\lambda:=\Tr B$, we finally obtain \eqref{thm:KRS}. 

\def\thesection{Appendix \Alph{section}} 
\section{} \label{proofs}
\def\thesection{\Alph{section}} 

For the proof of Lemma \ref{cor:error bound}, we need the following Lemma, which is essentially a restatement of inequality (44) in \cite{HN}. For readers' convenience, we give a detailed proof here.

\begin{lemma}\label{lemma:error bound}
For any $M\in\N$, any $c>0$, any $p$ finitely supported probability distribution on $\X$ and any $\pi:\,\supp p\to\B(\hil)$ such that $0\le\pi(x)\le I,\,x\in\supp p$, there exists an $M$-code $(M,\vfi,E)$ such that 
\begin{equation}\label{errprobupper}
P_{e}(M,\vfi,E)\le (1+c)\sum_x \prob{x} \Tr\channel{x}(I-\pi(x))+(2+c+1/c)(M-1)\sum_x \prob{x} \Tr E_p( W)\pi(x).
\end{equation}
\end{lemma}
\begin{proof}
For each $\vecc{x}\in\X^M$,  
define an $M$-code $(M,\vfi(\vecc{x}),E(\vecc{x}))$ by
\begin{equation*}
\vfi_k(\vecc{x}):=x_k,\ds\ds E_k(\vecc{x}):=\left[A_k(\vecc{x})+B_k(\vecc{x})\right]^{-\half}\,A_k(\vecc{x})\,\left[ A_k(\vecc{x})+B_k(\vecc{x})\right]^{-\half},\ds\ds k=1,\ldots,M,
\end{equation*}
where
\begin{equation*}
A_k(\vecc{x}):=\pi(x_k),\ds\ds\ds
B_k(\vecc{x}):=\sum\nolimits_{l\ne k}\pi(x_l),\ds\ds\ds k=1,\ldots,M.
\end{equation*}
Lemma 2 in \cite{HN} tells that $I-(A+B)^{-\half}A(A+B)^{-\half}\le (1+c)(I-A)+(2+c+1/c)B$ for any operators $0\le A\le I,\,0\le B$ on some Hilbert space $\hil$ and any number $c>0$. Applying it to $A=A_k(\vecc{x})$ and $B=B_k(\vecc{x})$, we get the bound
\begin{equation*}
I-E_k(\vecc{x})\le (1+c)(I-\pi(x_k))+(2+c+1/c) B_k(\vecc{x}),\ds\ds\ds c>0,
\end{equation*}
by which we get the following upper bound on the average error probability:
\begin{eqnarray}
P_{e}(\vecc{x})&:=&P_{e}(M,\vfi(\vecc{x}),E(\vecc{x}))=
\frac{1}{M}\sum_{k=1}^M \Tr \channel{x_k}\bz I-E_k(\vecc{x})\jz\nonumber\\
&\le& 
\frac{1+c}{M}\sum_{k=1}^M \Tr \channel{x_k}\bz I-\pi(x_k)\jz+\frac{2+c+1/c}{M}\sum_{k=1}^M \Tr\channel{x_k} B_k(\vecc{x}).\label{errprob2}
\end{eqnarray}
Note that for each $k$, $\vecc{x}\mapsto \channel{x_k}$ and $\vecc{x}\mapsto B_k(\vecc{x})$ are independent random variables on $\X^M$ with respect to any product measure on $\X^M$. Hence, taking the expectation value 
of both sides of the inequality in \eqref{errprob2} with respect to the product measure $p^{\otimes M}$ yields that $E_{p^{\otimes M}}P_e$ is upper bounded by the right-hand side of \eqref{errprobupper}.
Therefore, there has to exist at least one $\vecc{x}\in\bz\supp p\jz^M$ for which 
$P_{e}(\vecc{x})$ is upper bounded by the right-hand side of \eqref{errprobupper}, from which the assertion follows.
\end{proof}

\bigskip

\textit{Proof of Lemma \ref{cor:error bound}}
For a function $\pi:\,\supp p\to\B(\hil)$ such that $0\le \pi(x)\le I,\s x\in\supp p$, define  $\Pi:=\sum_{x\in\X}\delta_x\otimes\pi(x)$. With this notation, the upper bound in \eqref{errprobupper} can be rewritten as
\begin{equation}\label{error bound02}
\Tr A(I-\Pi)+\Tr B\Pi,\ds\ds
A:=(1+c)R_p,\ds\ds B:=(2+c+1/c)(M-1)Q_p,
\end{equation}
% \begin{align}
% &(1+c)\Tr R_p(I-\Pi)+(2+c+1/c)(M-1)\Tr Q_p\Pi\nonumber\\
% &\ds =1+c-\Tr\left[(1+c)R_p-(2+c+1/c)(M-1)Q_p\right]\Pi,\label{error bound02}
% \end{align}
which is minimized over all possible choices of $\Pi$ at the 
\ki{Holevo-Helstr\"om test}
% \begin{align*}
% \Pi^*&:=\left\{(1+c)R_p-(2+c+1/c)(M-1)Q_p> 0\right\}\\
% &=\sum_x\delta_x\otimes\{(1+c)W_x-(2+c+1/c)(M-1)E_p(W)>0\},
% \end{align*}
\begin{equation*}
\{A-B>0\}=\sum_x\delta_x\otimes\{(1+c)W_x-(2+c+1/c)(M-1)E_p(W)>0\}.
\end{equation*}
(Here we use the notation $\{X> 0\}$ to denote the spectral projection corresponding to the positive part of the spectrum of a self-adjoint operator $X$.)
Choosing therefore $\pi(x):=\{(1+c)W_x-(2+c+1/c)(M-1)E_p(W)>0\}$ in Lemma \ref{lemma:error bound}, we get the existence of an $M$-code for which the average error probability is upper bounded by the value of \eqref{error bound02} at $\Pi^*$, which is easily seen to be
$\half\Tr(A+B)-\half\Tr|A-B|$.
%  with
% \begin{equation*}
% A:=(1+c)R_p,\ds\ds\ds B:=(2+c+1/c)(M-1)Q_p.
% \end{equation*}
By Theorem 1 in \cite{Aud}, 
\begin{equation*}
\half\Tr(A+B)-\half\Tr|A-B|\le\Tr A^tB^{1-t}
\end{equation*}
for any two positive semidefinite operators $A$ and $B$ on some Hilbert space $\hil$ and any $t\in [0,1]$. Applying it to the above choice of $A$ and $B$, we finally get the assertion of the Lemma.
\hfill\qed

\def\thesection{Appendix \Alph{section}} 
\section{} \label{classical examples}

\def\thesection{\Alph{section}} 

\begin{ex} (\textbf{Classical state discrimination})\label{classical}

\noindent
Let $\rho_1,\ldots,\rho_M$ be commuting states on a Hilbert space $\hil$. Then, there exists a basis $e_1,\ldots,e_d$ of $\hil$ in which all the states are diagonal and hence they can be represented as functions on $\X:=\{1,\ldots,d\}$ in an obvious way. Let $\mathcal{E}(A):=\sum_{k=1}^d \inner{e_k}{Ae_k}\pr{e_k},\,A\in\B(\hil)$. If we use the POVM $E_1,\ldots,E_M$ to discriminate the states then the corresponding succes probability is 
\begin{equation*}
P_s(E)=\frac{1}{M}\sum_{i=1}^M\Tr\rho_i E_i=\frac{1}{M}\sum_{i=1}^M\Tr\rho_i \mathcal{E}(E_i),
\end{equation*}
and hence we can assume without loss of generality that the POVM operators are also functions on $\X$.
The succes probability is then
\begin{eqnarray*}
P_s(E)&=&\frac{1}{M}\sum_{i=1}^M\sum_{k=1}^d \rho_i(k)E_i(k)
=
\frac{1}{M}\sum_{k=1}^d \sum_{i=1}^M \rho_i(k) E_i(k)
\le
\frac{1}{M}\sum_{k=1}^d \sum_{i=1}^M m(k) E_i(k)\\
&\le&
 \frac{1}{M}\sum_{k=1}^d  m(k)  \sum_{i=1}^M E_i(k)
\le
\frac{1}{M}\sum_{k=1}^d m(k),
\end{eqnarray*}
where $m(k):=\max_{1\le i\le M}\rho_i(k),\,k=1,\ldots,d$. 
We say that $E_1,\ldots,E_M$ is a maximum likelihood measurement if $\sum_{i=1}^M E_i=I$ and 
$E_i(k)=0$ when $\rho_i(k)\ne m(k)$. It is easy to see that all the above inequalities hold with equality for any maximum likelihood measurement. Therefore, the optimal succes probability of discriminating the states  $\rho_1,\ldots,\rho_M$ is 
\begin{equation*}
P_s^*=\frac{1}{M}\sum_{k=1}^d m(k),
\end{equation*}
and hence, by \eqref{thm:KRS}, the max-relative entropy radius of $\{\rho_1,\ldots,\rho_M\}$ is 
\begin{equation}\label{classical maxradius}
R_{\max}\bz \{\rho_1,\ldots,\rho_M\}\jz=\log\sum_{k=1}^d m(k).
\end{equation}
\end{ex}

\begin{ex} (\textbf{Capacities of the classical depolarizing channel}) \label{classical depolarizing}

\noindent
We define the classical depolarizing channel $W^{d,\alpha}$ with parameters $d\in\N$ and $\alpha\in (0,1)$ the following way. Let $\X:=\{1,\ldots,d\}$ and $\hil:=\iC^d$ and let $\delta_k:=\pr{\egy_{\{k\}}}$, where $\egy_{\{1\}},\ldots,\egy_{\{d\}}$ is the standard basis of $\iC^d$. The action of the channel is
\begin{equation*}
W^{(d,\alpha)}:\,k\mapsto \alpha\delta_k+(1-\alpha)\frac{1}{d}I,\ds\ds\ds 1\le k\le d.
\end{equation*}
Note that the outputs of the channel are diagonal in the standard basis of $\iC^d$ and hence we will identify them with functions on $\X$ in an obvious way. 
Consider an $M$-code $(M,\vfi,E)$ and define 
\begin{equation*}
m(k):=\max_{1\le i\le M} W^{(d,\alpha)}_{\vfi(i)}(k)=\frac{1-\alpha}{d}+\alpha\max_{1\le i\le M}\delta_{\vfi(i)}(k),\ds\ds\ds 1\le k\le d.
\end{equation*}
Note that 
\begin{eqnarray*}
\sum_{k=1}^d m(k)&=&
1-\alpha+\alpha\sum_{k=1}^d\max_{1\le i\le M}\delta_{\vfi(i)}(k)\le
1-\alpha+\alpha\sum_{k=1}^d\sum_{i=1}^M\delta_{\vfi(i)}(k)\\
&=&
1-\alpha+\alpha\sum_{i=1}^M\sum_{k=1}^d\delta_{\vfi(i)}(k)=
1-\alpha+\alpha M.
\end{eqnarray*}
Hence, by Example \ref{classical}, the success probability of any code is upper bounded as
\begin{equation*}
P_s(M,\vfi,E)\le\frac{1-\alpha}{M}+\alpha.
\end{equation*}
If $\ep\in [0,1/2)$ then there exists an $\alpha\in(0,1)$ such that 
$1-\ep> \frac{1-\alpha}{2}+\alpha$ and hence, by the above bound, there exists no code $(M,\vfi,E)$ with $M>1$ and $P_e(M,\vfi,E)\le \ep$. Therefore, $C_{\ep}\bz W^{(d,\alpha)}\jz=0$. 

On the other hand, the Holevo capacity is 
\begin{eqnarray*}
\chi^*\bz W^{(d,\alpha)}\jz&=&
\sup_{p\in\M_f(\X)}\left\{S\bz \sum_{k=1}^d p_k W^{(d,\alpha)}_k\jz-\sum_{k=1}^d p_k S\bz W^{(d,\alpha)}_k\jz\right\}\\
&=&\sup_{p\in\M_f(\X)}\left\{S\bz \sum_{k=1}^d p_k W^{(d,\alpha)}_k\jz\right\}-S\bz W^{(d,\alpha)}_1\jz,
\end{eqnarray*}
where we have used that $W^{(d,\alpha)}_k$ is a permutation of $W^{(d,\alpha)}_1$ and hence their entropies are equal. The first term is clearly upper bounded by $S((1/d)I)=\log d$, which can actually be reached by choosing $p$ to be the uniform distribution on $\{1,\ldots,d\}$. Therefore,
\begin{equation*}
\chi^*\bz W^{(d,\alpha)}\jz=\log d-S\bz W^{(d,\alpha)}_1\jz,
\end{equation*}
and a straightforward computation yields
\begin{equation*}
\chi^*\bz W^{(d,\alpha)}\jz=\frac{1+(d-1)\alpha}{d}\log\bz 1+(d-1)\alpha\jz+\frac{d-1}{d}(1-\alpha)\log(1-\alpha),
\end{equation*}
which scales as $\log d$ as $d$ tends to infinity, for any parameter $\alpha\in (0,1)$.

Finally, by \eqref{classical maxradius}, the max-relative entropy radius of $\ran W^{d,\alpha}$ is given by
\begin{equation*}
R_{\max}\bz\ran W^{d,\alpha}\jz=\log\sum_{k=1}^d\max_{1\le i\le d}W^{d,\alpha}_i(k)=\log(1+(d-1)\alpha).
\end{equation*}
Note that
\begin{equation*}
R_{\max}\bz\ran W^{d,\alpha}\jz-\chi^*\bz W^{(d,\alpha)}\jz=\frac{(d-1)(1-\alpha)}{d}\log\frac{1-\alpha+d\alpha}{1-\alpha}>0.
\end{equation*}
Since the Holevo capacity is equal to the relative entropy radius $R\bz\ran W^{d,\alpha}\jz$ by \cite{OPW}, this shows that 
\begin{equation*}
R_{\max}\bz\ran W^{d,\alpha}\jz>R\bz\ran W^{d,\alpha}\jz.
\end{equation*}
\end{ex}

\end{document}